\documentclass[10pt]{wiley2sp}
\usepackage{url}
\usepackage{helvet}
\usepackage{graphicx}
\usepackage{amsmath}

\def\pages{}        
\def\thevol{}          
\def\thenumber{}       
\def\theyear{2005}      
\def\thedoi{}   

\def\PRL{{Phys. Rev. Lett.}}
\def\PRA{{Phys. Rev.} A }

\newcommand{\be}{\begin{equation}}
\newcommand{\ee}{\end{equation}}
\newcommand{\bea}{\vspace{0.25cm}\begin{eqnarray}}
\newcommand{\eea}{\end{eqnarray}}

\title{Numerical systematical study of atmospheric effects on Earth-Space QKD}
\author{N. Antonietti,\inst{1}$^{,\ast}$
        M. Mondin,\inst{1}
        G. Catastini,\inst{2}
        G. Brida,\inst{3}
        M. Genovese\inst{3}
}

\institute{Dipartimento di Elettronica, Politecnico di Torino, Corso
Duca degli Abruzzi 24, 10129, Torino - Italy
         \and
         Alcatel Alenia Space, Corso marche 41, 10146, Torino - Italy
         \and
         I.N.RI.M., Strada delle Cacce 91, 10135, Torino - Italy.}

\mail{e-mail: nantoni@calstatela.edu}

\begin{document}

\abstract{In this paper, the authors compare the security bounds for
different quantum communication protocols with the numerically
evaluated losses in the transmission channel, due to the interaction
between the atmosphere and the photon, that is the information
carrier. The analysis is carried out using a free-source library
which can solve the radiative transfer equation for a parallel plane
atmosphere.}


\maketitle
\sloppy

\section{Introduction}

In the last ten years a new discipline called quantum information
and devoted to codification, elaboration and transmission of
information by exploiting the specific properties of quantum systems
has been widely studied and tested.\\
The information is said to be quantum when it is encoded in quantum
system (for instance the spin of a particle, the polarization or
phase of a photon). A secure cryptosystem can be achieved if one
encodes information in a quantum system. To be more precise, by
exploiting the properties of quantum systems, Alice and Bob can
share secret keys which can then be used in standard secret-key
protocols \cite{gis,w,5,6,7}. The most natural carrier of quantum
information is the photon. In fact it travels with the speed of
light, has a very limited interaction with the environment and
allows to encode information in several degrees of freedom, such as
polarization, phase or energy. All the experimental realizations of
quantum cryptographic protocols (more properly Quantum Key
Distribution protocols, QKD), since the first one at IBM in 1989
(published in 1992 \cite{5}) used single photons as quantum bits
(qubits). Afterwards, many research teams have made quantum key
distribution using different protocols and different physical
implementations \cite{gis}.

Following these experimental verifications, the research on QKD has
been then addressed toward the realization of long distance
communications \cite{lon}, the implementation of protocols suited
for commercial purposes \cite{com}, the study of eavesdropping in
realistic conditions \cite{eav}, protocols in higher dimensions
and/or many particles \cite{qud,las2}, ...

 Nowadays the frontier of QKD is the realization of a
Earth-Space (Space-Earth) or a Space-Space quantum communication
channel \cite{zei,rar,giap,nos}. A communication channel, in this
case the atmosphere, is said to be quantum when quantum information
is transmitted through it and we call this process quantum
communication.

This realization would be of utmost relevance both for quantum
key distribution \cite{gis}, since it would allow intercontinental
quantum transmissions, and for studies concerning foundations of
quantum mechanics \cite{zei,mg}. Thus, preliminary feasibility
studies have been performed showing its practical realizability.

In little more details, in ref. \cite{rar} a BB84 scheme was studied
for Earth-Space communication. By considering gaussian optics, a 15
dB loss was attributed to diffraction, whilst aerosol loss was
considered of secondary relevance (0.04-0.06 dB) for  transmission
with clear sky and from high elevation above sea. Altogether
atmospheric losses were estimated to be about 2-5 dB. On the other
hand, the security level for quantum transmission (the amount of
losses which the communication is still safe with) was estimated to
be 40 dB (from estimated background and detectors dark counts),
lowering to 10 dB if the eavesdropper (conventionally dubbed Eve)
had technologies for intercepting selectively a possible
multi-photon component.

In another study, ref. \cite{zei}, a 6.5 dB loss was estimated by
considering optics and finite quantum efficiency of detectors \cite{las1}. The
limit for secure quantum transmission was estimated to be at 60 dB
loss. Finally, here atmospheric losses were estimated to be around 1
dB.\\
Effectively, daylight and open space transmission at a distance of
10 km \cite{10,rz} was achieved and, very recently, an European
collaboration has achieved preliminary results on a quantum channel
(at Canary islands) at a distance up to 144 km \cite{tom}.

All these theoretical and experimental studies guarantee the
feasibility of a ground-space channel. Nevertheless, the analysis of
atmospheric effects is rather incomplete and is far from considering
various realistic atmospheric situations that could be met during a
real transmission. Even in the very general review of ref.
\cite{gil}, only few results are presented, and with small detail.

Thus, a detailed analysis of atmospheric effects in various
realistic situations would be of the utmost relevance. In particular
one should consider both static atmosphere effects (as absorbtion,
scattering and emissions) and turbulent atmospheric effects (as
wandering). Even if the latter represent the main contribution to
losses, more than reflection, in perspective they can be coped with technological
solutions (as adaptative optics), whilst absorption (even if
smaller) will represent the final limit of this kind of
communication. Errors due to electronic and optical imperfections must be taken into account separately.

Purpose of this paper is to describe a work that addresses a
precise characterization of static atmospheric effects on the
quantum communication process.

We want to determine the losses in the quantum information carriers
(photons) and, in this work of ours, we consider lost any photon
that has interacted with the atmosphere. Then we compare the
estimated losses with the security loss upper bounds mentioned
before. After estimating the dependence of a secure transmission in
different meteorological situation, for example, it would then be
possible to evaluate the average available time for a secure
communication for a certain ground station by the statistical
meteorological conditions of the station itself.

To investigate this topic we have used a free source library for
radiative transfer calculations named libRadtran\cite{libradtran}.
This library can solve the radiative transfer equations, set some
input parameters and exploit the HITRAN\cite{hitran} database, which
is a high-resolution atmospheric parameters database (for example
there are the atmospherical components cross sections for different
wavelengths).\\
In our simulations we can determine what part of the irradiance of a
source at the top of the atmosphere can reach the ground without
interactions with the atmosphere. The effects on photon polarization
can be estimated to be small albeit not completely negligible
(depolarization being of the order of $3.5 \%$ by including single
forward Rayleigh scattering only \footnote{forward Mie scattering
depolarization being anyway negligible.}): their precise evaluation
is now in progress.

Various parameters can influence the atmospheric effects on the
photon transmission, as, for instance, aerosols, pressure,
temperature, air density, precipitations, cloud composition,
humidity, chemical components. As a first step in order to evaluate
their relevance in various meteorological conditions, here we
present some preliminary results obtained by varying some of them in
realistic intervals.

\section{Numerical results}
\label{sec:real_experiments}

In \cite{chand}, a \textit{pencil of radiation}\ is defined in terms
of the \textit{specific intensity}\ $u_{\nu}$, as the amount of
energy $dE_{\nu}$, in a specified frequency interval $(\nu ,\nu +
d\nu)$, which is transported across an element of area $d\sigma$,
along directions confined to an element of solid angle $d\omega$,
during a time $dt$, according to the following formula:

\begin{equation}
    dE_{\nu} = u_{\nu} \cos \theta d\nu d\sigma d\omega dt,
\end{equation} where $\theta$\ is the angle between the direction
of the incoming radiation and the outward direction normal to
$d\sigma$.\\
We can consider a photon traveling through the atmosphere, as a
pencil of radiation whose radiant energy is discrete and ideally as
a monochromatic radiation (they have actually a bandwidth, although
very narrow).\\
Thus, in this sense, the photon can be considered with a classical
treatment, undergoing absorption and scattering.\\
Let's then consider an atmosphere modeled as stratified in parallel
planes. In such an atmosphere, all the physical properties
(temperature, pressure, relative humidity...) and the chemical
component density are invariant over a plane. That's a very common
model for the atmosphere that is, indeed very often in nature,
structured as plane parallel \cite{chand}.\\
When a pencil of radiation traverses a medium of density $\rho$\ and
thickness $ds$, it's weakened according to the following equation,
$du_{\nu} = -k_{\nu} \rho u_{\nu} ds$, where $k_{\nu}$\ is called
the mass absorption coefficient.\\
The optical depth at the height $z$ from the Earth surface is $\tau
= \int_z^{\infty}k \rho dz$.

The transfer of monochromatic radiation through a parallel plane
atmosphere is described by the radiative transfer equation
\cite{chand}:

\begin{equation}\label{rte}
    \mu \frac{du_{\nu}(\tau_{\nu}, \mu, \phi)}{d\tau}=u_{\nu}(\tau_{\nu}, \mu,
    \phi) - S_{\nu}(\tau_{\nu}, \mu, \phi)
\end{equation} where $\mu=\cos \theta$, $u_{\nu}(\tau_{\nu}, \mu, \phi)$\ is the
specific intensity along the direction $(\mu ,\phi )$, at optical
depth $\tau _{\nu}$.\\

 $S$\ is the source function:

\begin{equation}\label{eq:soft_sourc_func}
\begin{aligned}
    & S_{\nu}(\tau _{\nu},\mu ,\phi) = \frac{\omega _{\nu}(\tau _{\nu})}{4\pi}
    \int_0^{2\pi} d\phi ' \int_{-1}^1 d\mu 'P_{\nu}(\tau _{\nu},\mu ,\phi; \mu ',\phi
    ')\\
    & \times u_{\nu}(\tau _{\nu},\mu' ,\phi') + Q_{\nu}(\tau _{\nu},\mu
    ,\phi),
\end{aligned}
\end{equation} $\omega_{\nu}(\tau_{\nu})= \int p(\cos \Theta)\frac{d\omega
'}{4\pi}$\ is the single scattering albedo and $p(\cos \Theta )$\ is
the phase function, that gives the rate at which energy is being
scattered into an element of solid angle $d\omega '$\ and in a
direction inclined at a direction $\Theta$\ to the direction of
incidence of a pencil of radiation on an element of mass $dm$. For
our investigations, a Henyey-Greenstein phase function is assumed.\\
$Q_{\nu}(\tau _{\nu},\mu ,\phi) = [1-\omega_{\nu}
(\tau_{\nu})B_{\nu}(T(\tau_{\nu}))]$\ is the amount of radiant
energy emitted by the atmosphere as thermal emission and
$B_{\nu}(T(\tau_{\nu}))$\ is the Planck function at frequency $\nu$\
and temperature $T$.\\

The discrete approximation to (\ref{rte}) can be written as
\cite{stamnes}:

\begin{equation}\label{rte_disc}
    \begin{aligned}
    & \mu_i \frac{du^m (\tau ,\mu_i )}{d\tau} = u^m (\tau ,\mu_i ) -
    \sum_{\substack{j=-N \\ j\neq 0}}^N w_jD^m (\tau ,\mu_i ,\mu_j ) \times \\
    & u^m (\tau ,\mu_j) - Q^m (\tau ,\mu_i) \hspace{3mm} (i=\pm 1 ,\ldots \pm N ).
    \end{aligned}
\end{equation} The phase function is expanded in a series of
Legendre polynomials ($D^m$) and the intensity in a Fourier cosine
series whose coefficient are $u^m$. $w_i$\ are the quadrature
weights of the series of Legendre polynomials. Equation (\ref{rte})
is replaced by 2N independent equations (whose unknown quantities
are the coefficients $u^m$) and the procedure is repeated for each
layer we have divided the atmosphere in. Equation (\ref{rte_disc})
is transformed in a system of $2N$-coupled ordinary differential
equations with constants coefficients which can be solved
numerically.

The solution of (\ref{rte_disc}) is reported in \cite{stamnes}. The
libRadtran library we are going to use solves the radiative transfer
equation (\ref{rte}) and gives, at the output, the amount of energy
which doesn't interact with the atmosphere, by means of the
algorithm presented in \cite{stamnes}.

\subsection{Different distributions of aerosols}
\label{subsec:aerosols}

In libRadtran, a database of aerosols distributions and their
optical properties can be found. It has been written according to
ref. \cite{shettle:1989}. There, four aerosols distributions for
four different environment conditions are described (rural,
maritime, urban, tropospheric).\\

For our first analysis, the atmospheric conditions (for a
plane-parallel atmosphere model) are selected to be in summer
season, at midlatitudes, according to ref. \cite{afgl:1986}; there,
the atmosphere is described with its pressure, temperature, air
density, relative humidity profiles, etc and the optical properties
are derived. The source irradiance, posted at the top of the
atmosphere, is chosen in accordance to ref.
\cite{kato:1999}; both databases are present in libRadtran.\\

For these conditions the direct downward irradiance (the amount of
radiation which doesn't experience any interaction with the
atmosphere) at the Earth surface with the source at the zenith and
at a zenith angle of either 50$^o$ and 80$^o$, outside the whole
atmosphere has been evaluated. Calculations are performed invoking a
correlated-k band parametrization by Kato et al. \cite{kato:1999}
and this choice will be maintained for the next calculations, unless
differently specified. We are mainly interested in the visible
wavelengths (actually in between about 700 nm and 900 nm) because
that range is not affected by strong absorption as UV and IR bands;
anyway, for completeness we extended our investigations in the
infrared band and actually the exact range is in between 256.3 nm
and 2638.5 nm (being some good transmission windows present here as
well); the extreme points in the wavelength range are determined by
the choice of the database \cite{kato:1999}.

\begin{figure}[h]
\begin{flushleft}
  \includegraphics[width=8.5cm]{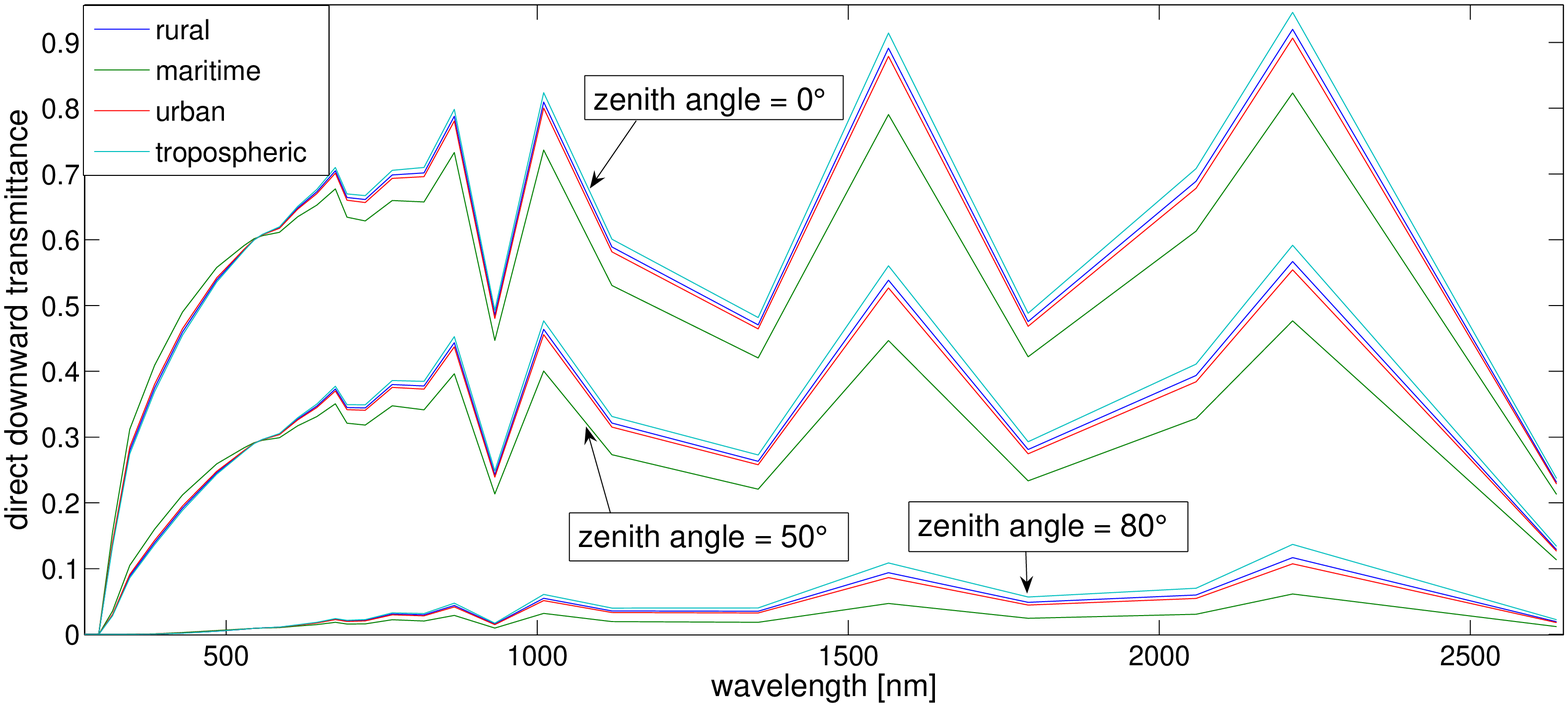}\\
  \caption{Direct downward transmittance (ratio between the direct downward
  irradiance at the Earth surface with respect to the source irradiance)
  vs wavelength in nm for 0$^o$, 50$^o$, 80$^o$ source zenith angles and different aerosols
  conditions, in the range between 256.3 nm and 2638.5 nm.
  The atmosphere is in summer conditions and at midlatitudes, according
  to ref. \cite{afgl:1986} and the source irradiance is chosen in accordance
  to ref. \cite{kato:1999}. The aerosol distributions are the rural,
  maritime, urban and tropospheric as described in ref. \cite{shettle:1989}. Here and in the following
   large scale figures the wave length resolution is kept poor in order not to compromise the
   readability:  the code allow a much
   more detailed resolution than can be exploited for the region of interest (and that will be used in some
   smaller scale figures).
   }
  \label{transmittance_shettle_aerosols}
\end{flushleft}
\end{figure}

In figure (\ref{transmittance_shettle_aerosols}) we show the direct
downward transmittance $T$ (ratio between the direct downward
irradiance at the Earth surface with respect to the source
irradiance). A first result of this analysis is that this quantity
is largely independent of the aerosol type. Moreover the behavior of
the transmittances for different aerosols conditions are largely
independent of the source zenith angle too.\\
Of course, the evaluation of $T$ next to the extreme source zenith
angle of 90$^o$\ is useless, because fairly no radiation reaches the
ground (besides the parallel planes model of atmosphere used in the
program cannot be extended beyond 80$^o$ from zenith).

It can be observed that the best range for communications is roughly
from 600 nm to almost 900 nm, but some window is present also in
infrared region (as 1,564 nm or 2,214 nm).
\begin{figure}[h]
\begin{flushleft}
  \includegraphics[width=8.5cm]{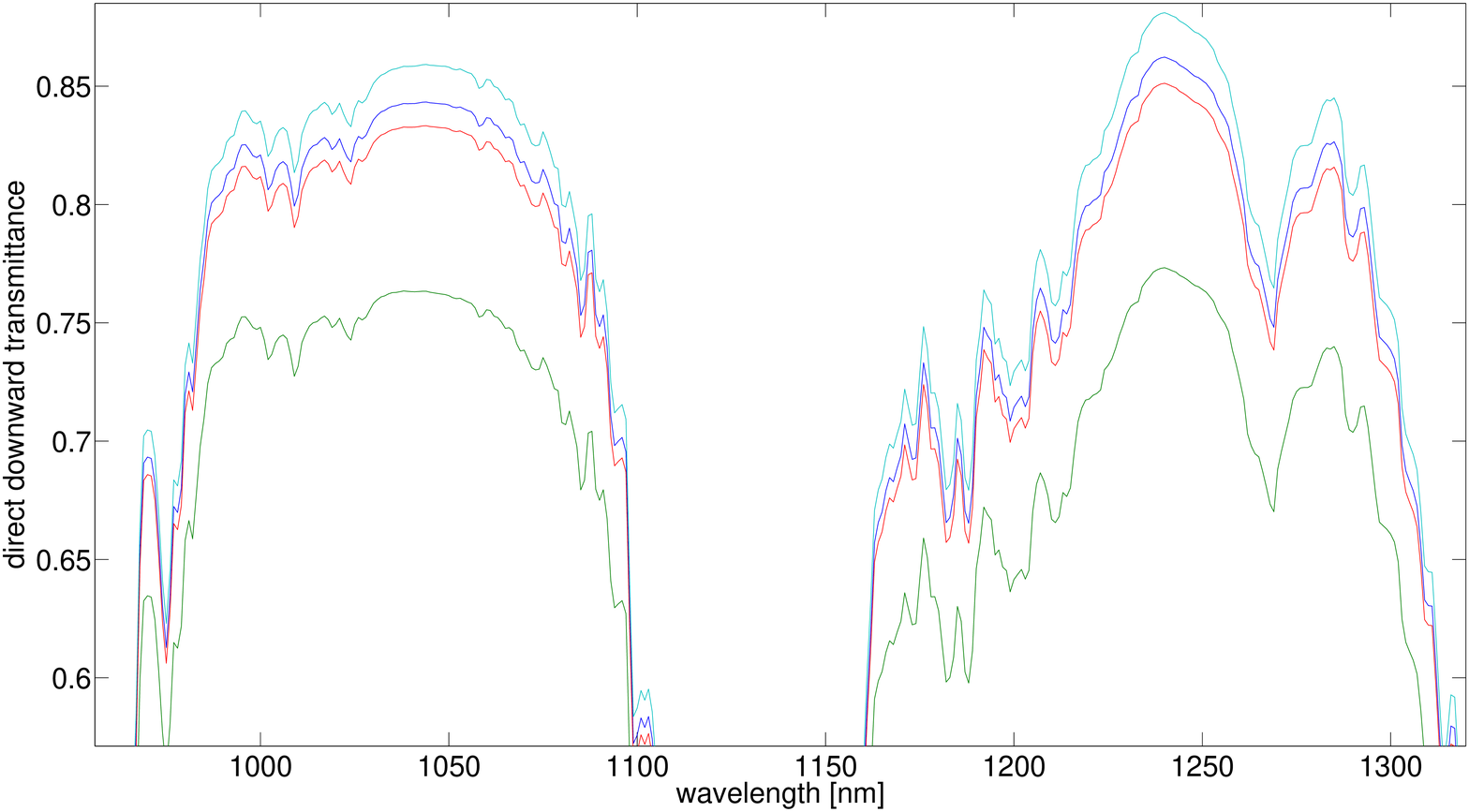}
   \includegraphics[width=8.5cm]{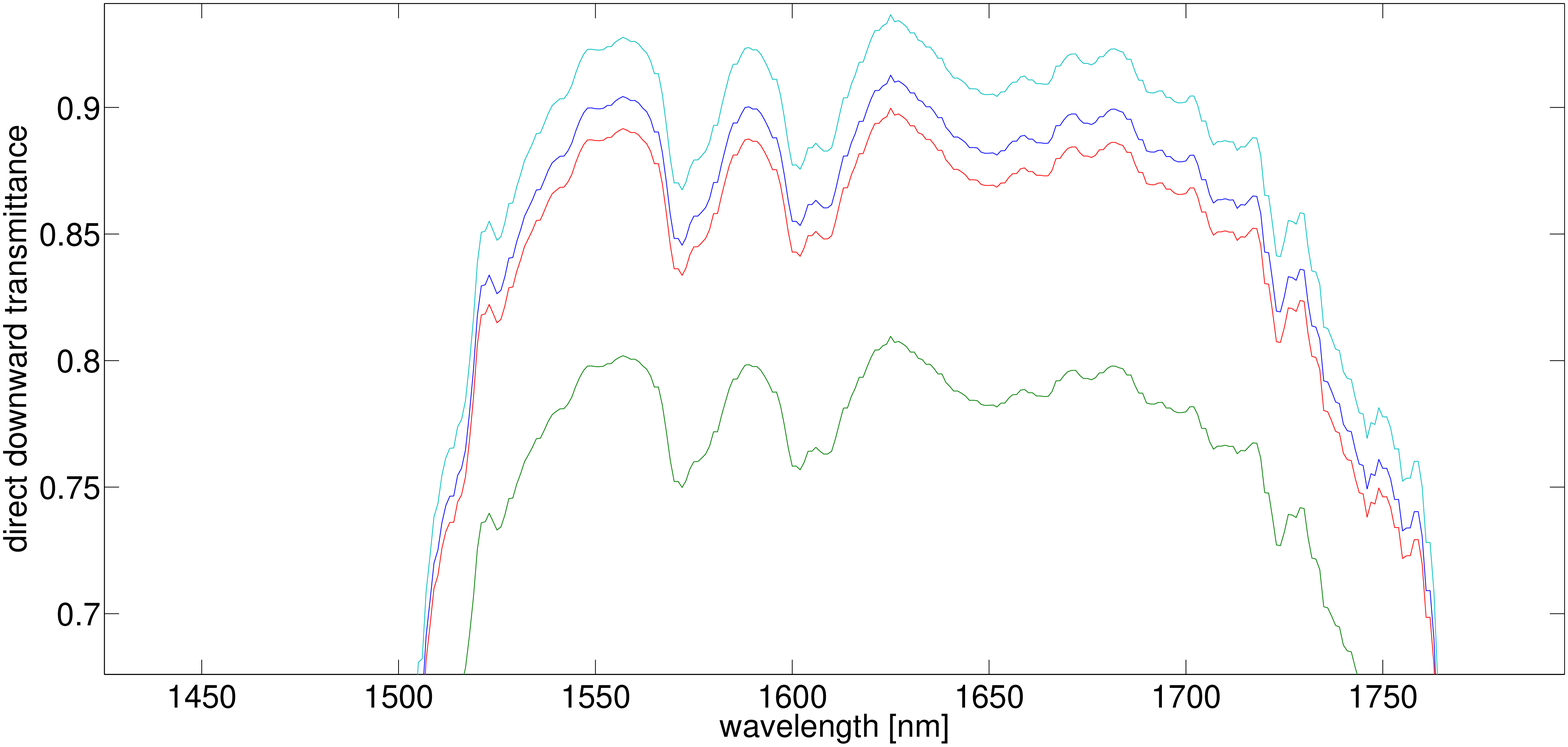}
\includegraphics[width=8.5cm]{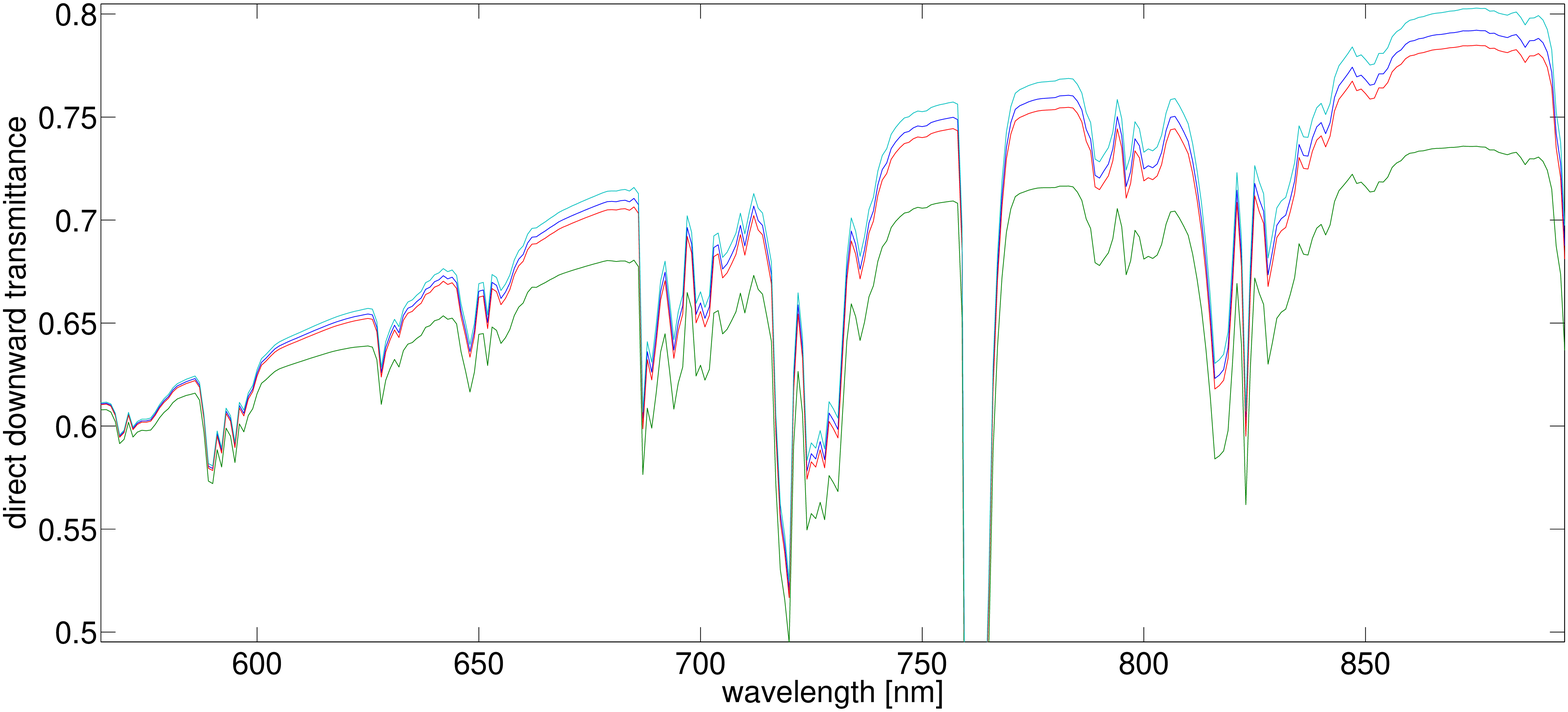}\\
  \caption{As in fig.1, but selecting three specific wave length windows
  with LOWTRAN resolution.
   }
  \label{aerdet}
\end{flushleft}
\end{figure}
A detail of absorptions in some more restricted wave-length windows
is shown in fig. (\ref{aerdet}), using the LOWTRAN atmospheric
database. Incidentally, the use of strong resolution in wave length
makes unreadable figures with a large scales and thus we only show
it in lower scale figures as (\ref{aerdet}); here one can appreciate
the details and clearly distinguish specific absorption lines that
should be avoided for communication. On the other hand large scale
figures show the rough general dependence of absorption, pointing
out the regions where a more detailed analysis can be interesting.

In the range 600-900 nm, for a source zenith angle of 0$^o$, the
fraction source light which gets across the atmosphere without any
interaction with the atmosphere is, excluding some specific
absorption lines (as, for instance, around 760 nm, that is due to the oxygen absorption line), about 70\%.\\
This means that a photon has a 70\% probability to get across the
model of atmosphere we have built without interacting with it.\\
We can present, as usual, the losses in dB ($l_{dB}$), from the
direct downward transmittance in percent ($T_{\%}$),

\begin{equation}
    l_{dB} = -10 \log_{10}(T_{\%})
\end{equation}

Then, in the visible window from 700 nm to 900 nm, the losses due to
atmospheric interaction are less than 4 dB if the source zenith
angle is 0$^o$\ and less than 20 dB if the source zenith angle is
80$^o$, but at
any rate lower than requested in \cite{zei} for establishing a
secure communication. Thus, one can infer that the transmission can
be carried on for almost the whole visibility range of a satellite,
when the atmosphere is in the conditions we have described before.

In the following picture (\ref{losses_aerosols} )the losses versus
the wavelength are depicted:

\begin{figure}
\begin{flushleft}
  \includegraphics[width=8.5cm]{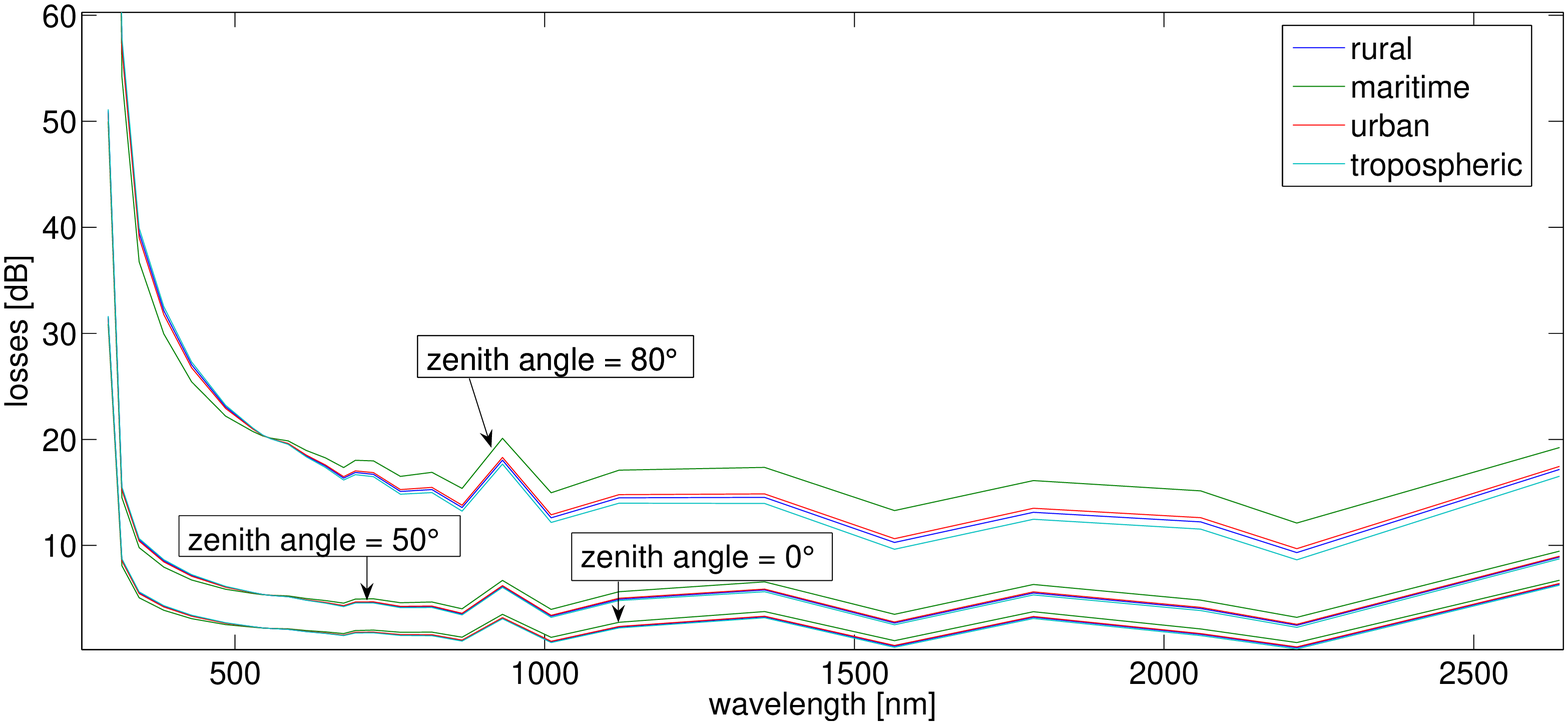}\\
  \caption{Losses in dB vs wavelength in nm for 0$^o$, 50$^o$, 80$^o$ source zenith angles and different aerosols
  conditions, in the range between 256.3 nm and 2638.5 nm.
  The atmosphere is in summer conditions and at midlatitudes, according
  to ref. \cite{afgl:1986} and the source irradiance is chosen in accordance
  to ref. \cite{kato:1999}. The aerosol distributions are the rural,
  maritime, urban and tropospheric as described in ref. \cite{shettle:1989}.
  The lowest and most constant losses are in the range from 700 nm to
  900 nm} \label{losses_aerosols}
\end{flushleft}
\end{figure}
When the zenith angle is 0$^o$, the losses are less than 60 dB for
wavelengths from 295.1 nm; when zenith angle is 50°$^o$ losses are
around 60 dB at wavelengths equal to 295.1 nm; finally, at zenith
angle equal to 80°$^o$, losses are less than 60 dB only starting
from 317.3 nm.

\subsection{Different temperature profile}
\label{subsec:temp_prof}

\begin{figure}[h]
\begin{flushleft}
  \includegraphics[width=8.5cm]{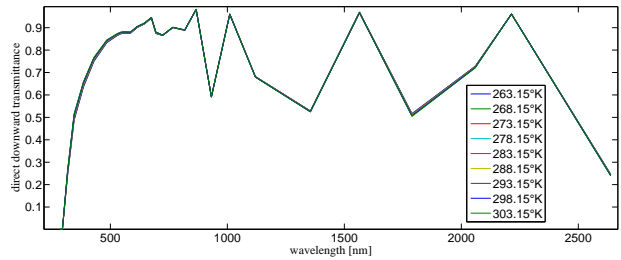}\\
  \caption{Direct downward transmittance (ratio between the direct downward
  irradiance at the Earth surface with respect to the source irradiance)
  vs wavelength in nm for 0$^o$ source zenith angle, in the range between 256.3 nm and 2638.5 nm.
  The atmosphere is in summer conditions and at midlatitudes, according
  to ref. \cite{afgl:1986}, but with a further modification: the surface
  temperature has been made vary from -10$^o$C up to 30$^o$C,
  with steps of 5$^o$C and the temperature has been decreased
  linearly with the altitude (for the first fifteen kilometers), up to
  the value it assumed in the unmodified atmosphere \cite{afgl:1986}.
  The source irradiance is chosen in accordance to ref. \cite{kato:1999}.
  This atmosphere is considered aerosol-free}
  \label{transmittance_temperature}
\end{flushleft}
\end{figure}

In the considered atmosphere database, the temperature decreases
fairly linearly from the ground level value $T_{0}$ up to 15 km,
where it assumes a given value $T_{15}$. $T_{0}$ is actually a free
parameter, and we have varied its value from -10$^o$C up to 30$^o$C,
with steps of 5$^o$C. The air density is modified according to the
perfect gas law. Above 15 km, the parameters have been left
unchanged, and no aerosols presence has been considered. We
evaluated the Transmittance for three different source zenith angles
(0$^o$, 50$^o$, 80$^o$). It turns out that the ground level
temperature doesn't affect the transmittance at all. Moreover, as in
the previous case, the behavior of the transmittances for different
level ground temperatures are largely independent of the
source zenith angle. This atmosphere and the following ones are aerosol-free\\
So, we depict in figure (\ref{transmittance_temperature}) the
results of the simulations for the source zenith angle equal to
0$^o$.

\begin{figure}
\begin{flushleft}
  \includegraphics[width=8.5cm]{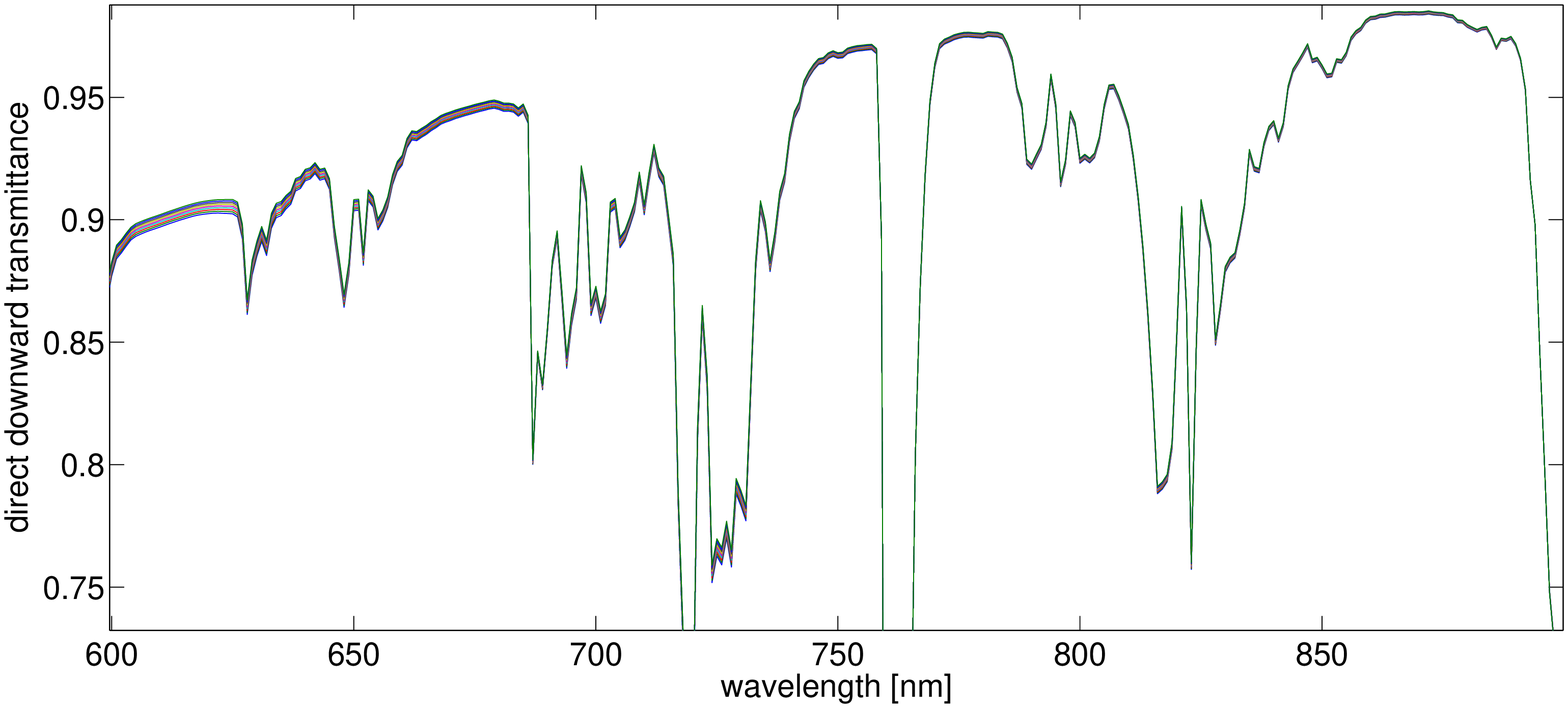}\\
  \caption{Detail of the figure \ref{transmittance_temperature}, from 600 nm to 900 nm}
  \label{transmittance_temperature_zoom}
\end{flushleft}
\end{figure}

Once again, it is possible to observe that the best range for the
communications is roughly from 700 nm to 900 nm as well. In fact, in
this case where no aerosol was included, at a solar zenith angle of
80$^o$ the losses are less than 14 dB. The detailed dependence,
according to the LOWTRAN atmospheric database, can be observed in
picture (\ref{transmittance_temperature_zoom}).

\begin{figure}
\begin{flushleft}
  \includegraphics[width=8.5cm]{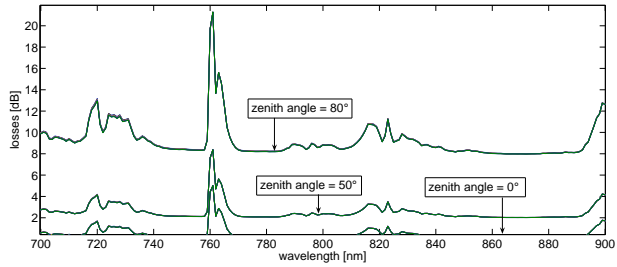}\\
  \caption{losses in dB vs wavelength in nm for 0$^o$, 50$^o$ and 80$^o$ source zenith angle, in the range between 700 nm and 900 nm.
  The atmosphere is in summer conditions and at midlatitudes, according
  to ref. \cite{afgl:1986}, but with a further modification: the surface
  temperature has been made vary from -10$^o$C up to 30$^o$C,
  with steps of 5$^o$C and the temperature has been decreased
  linearly with the altitude (for the first fifteen kilometers), up to
  the value it assumed in the unmodified atmosphere \cite{afgl:1986}.
  The source irradiance is chosen in accordance to ref. \cite{kato:1999}.
  This atmosphere is considered aerosol-free.}
  \label{losses_temperature}
\end{flushleft}
\end{figure}

As for the previous analysis, when the zenith angle is 0$^o$, the
losses are less than 60 dB for wavelengths from 295.1 nm; when
zenith angle is 50$^o$, losses are around 60 dB at wavelengths equal
to 295.1 nm and then decrease as wavelengths increase; finally, at
zenith angle equal to 80$^o$, losses are less than 60 dB only
starting from 317.3 nm.\\
In this case, the range from 700 nm to 900 nm is the actual minimum
absorbtion range in the visible wavelengths and in the picture
(\ref{losses_temperature}), it is seen in detail (hereafter, all the
detailed figures are imlied to be evaluated with the LOWTRAN
database).

\subsection{Different humidity profiles}
\label{subsec:hum_prof}

\begin{figure}[h]
\begin{flushleft}
  \includegraphics[width=8.5cm]{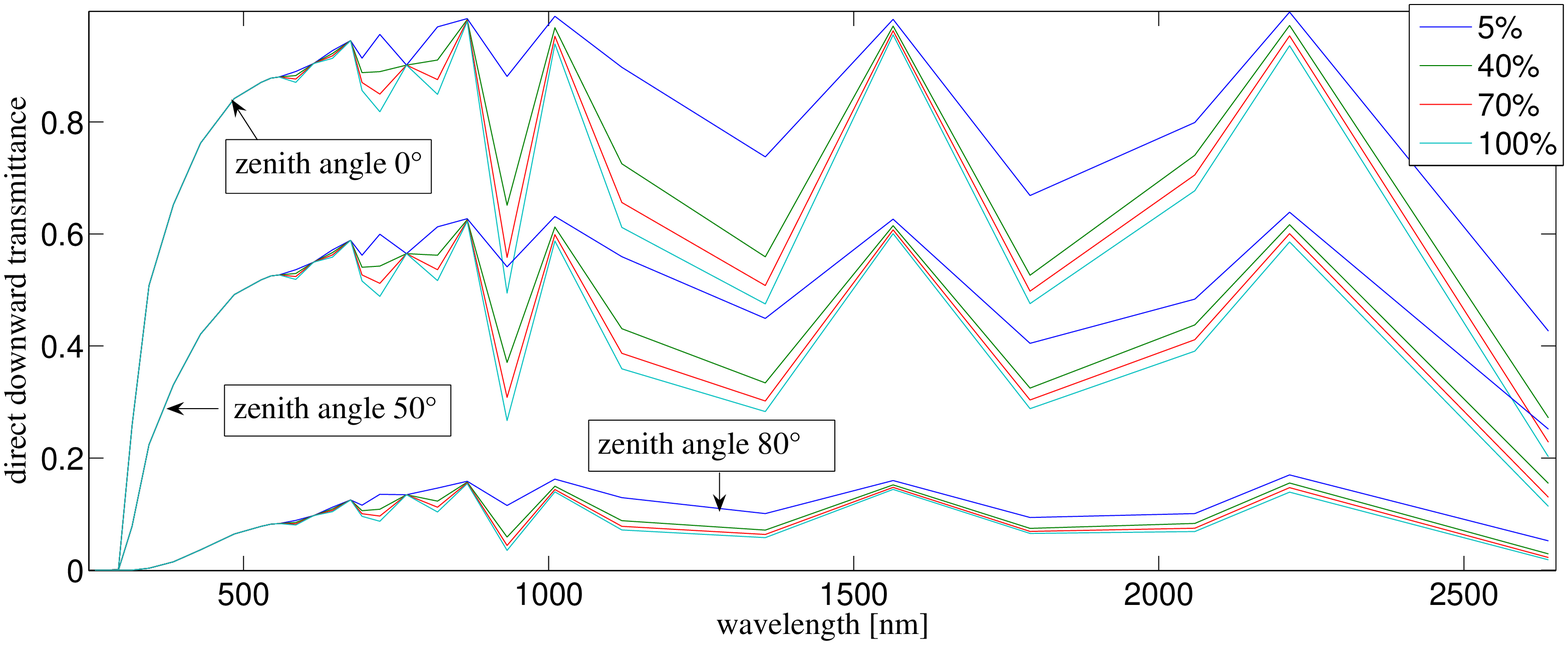}\\
  \caption{Direct downward transmittance (ratio between the direct downward
  irradiance at the Earth surface with respect to the source irradiance)
  vs wavelength in nm for 0$^o$, 50$^o$, 80$^o$ source zenith angles, in the range
  between 256.3 nm and 2638.5 nm.
  The atmosphere is in summer conditions and at midlatitudes, according
  to ref. \cite{afgl:1986}, but with a further modification: the relative
  humidity has been set constant along the first 15 kilometers of the
  atmosphere, with values of 5\%, 40\%, 70\% and 100\% values.
  The source irradiance is chosen in accordance to ref. \cite{kato:1999}.
  This atmosphere is considered aerosol-free}
  \label{transmittance_humidity}
\end{flushleft}
\end{figure}

In order to observe the effect of humidity, a further modification
has been added to the atmospheric conditions of ref.
\cite{afgl:1986}. This time, the relative humidity has been set in
different times as a constant value along the first 15 km of the
atmosphere. The values are 5\% and from 10\% to 100\% with steps of
10\%. No aerosols have been considered in this configuration. In
figure (\ref{transmittance_humidity}), the evaluated direct downward
transmittance can be observed for source zenith angles of 0$^o$,
50$^o$ and 80$^o$.

\begin{figure}[h]
\begin{flushleft}
  \includegraphics[width=8.5cm]{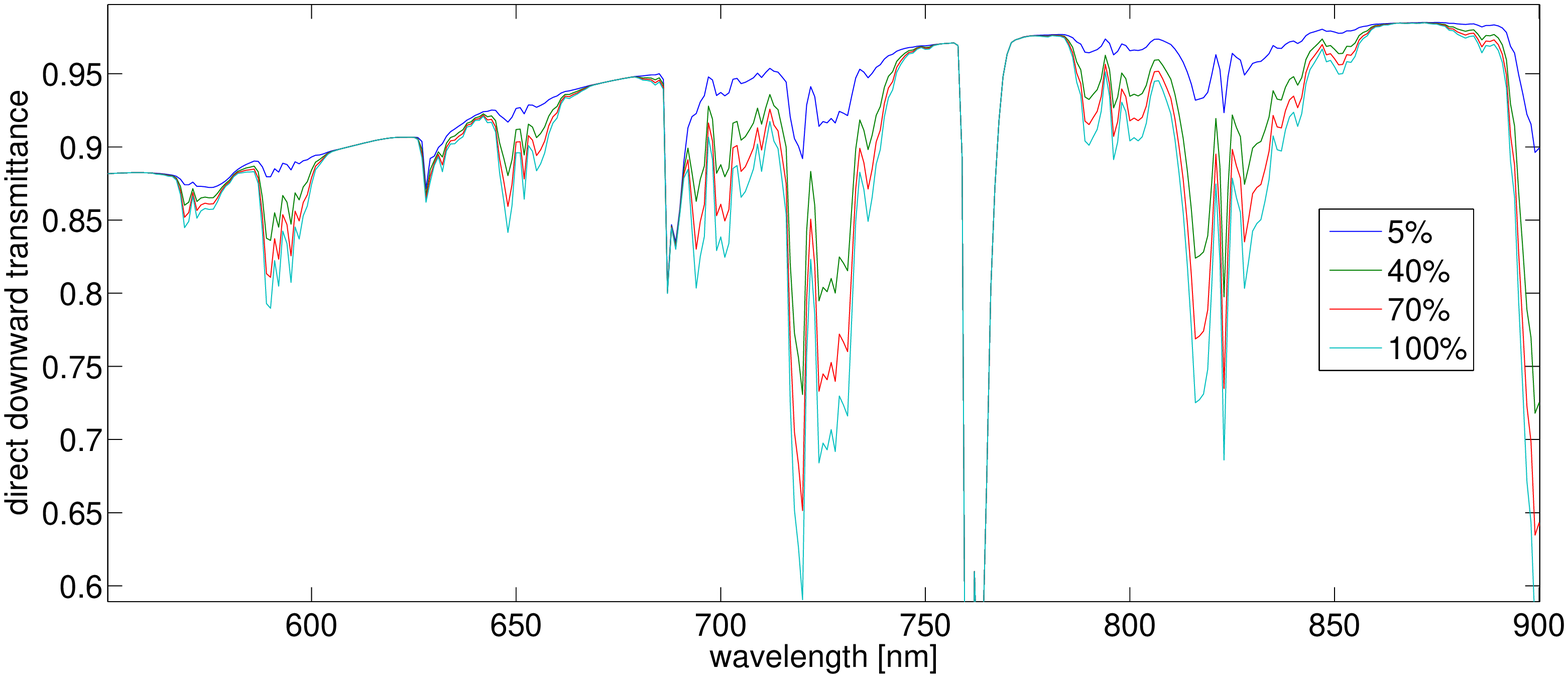}\\
  \caption{Detail of the figure \ref{transmittance_humidity}, from 550 nm to 900 nm and a 0$^o$ zenith angle}
  \label{transmittance_humidity_zoom}
\end{flushleft}
\end{figure}

Also in this scenario, the most advantageous range for communication
is from 700 nm to 900 nm, where losses are less than 10 dB, but in
this case we can observe some differences among different humidity
conditions. As it can be observed from figure
(\ref{transmittance_humidity}) and, in more details, in figure
(\ref{transmittance_humidity_zoom}), the absorption in the range
between 800 nm and 1000 nm is water vapor dependent and is strongly
affected by its presence.

\begin{figure}
\begin{flushleft}
  \includegraphics[width=8.5cm]{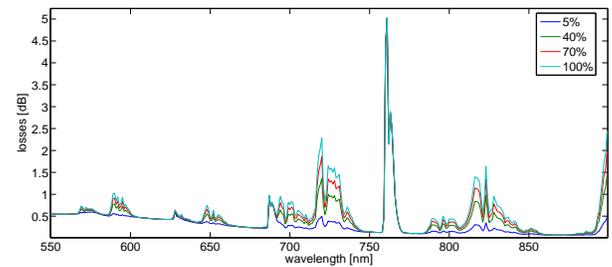}\\
  \caption{Losses in dB vs wavelength in nm for 0$^o$ source zenith angle, in the range
  between 550 nm and 900 nm.
  The atmosphere is in summer conditions and at midlatitudes, according
  to ref. \cite{afgl:1986}, but with a further modification: the relative
  humidity has been set constant along the first 15 kilometers of the
  atmosphere, with values of 5\%, 40\%, 70\% and 100\% values.
  The source irradiance is chosen in accordance to ref. \cite{kato:1999}.
  This atmosphere is considered aerosol-free}
  \label{losses_humidity}
\end{flushleft}
\end{figure}

The losses in dB are depicted in figure (\ref{losses_humidity}),

When the zenith angle is 0$^o$, the losses are less than 60 dB for
wavelengths from 295.1 nm; when zenith angle is 50°$^o$ losses are
around 60 dB at wavelengths equal to 295.1 nm; finally, at zenith
angle equal to 80°$^o$, losses are less than 60 dB only starting
from 317.3 nm.\\
There are minima outside the range from 700 nm to 900 nm but the
former is the most stable.

\subsection{Presence of clouds}
\label{subsec:cloud}

\begin{figure}[h]
\begin{flushleft}
  \includegraphics[width=8.5cm]{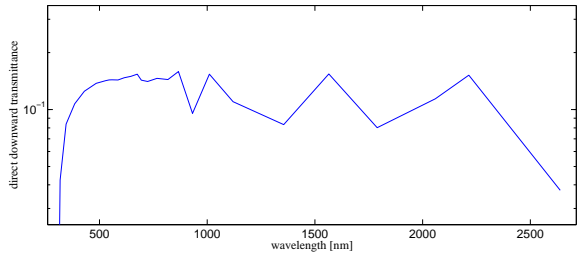}\\
  \caption{Direct downward transmittance (ratio between the direct downward
  irradiance at the Earth surface with respect to the source irradiance)
  vs wavelength in nm for 0$^o$ source zenith angle, in the range
  between 256.3 nm and 2638.59 nm.
  The atmosphere is in summer conditions and at midlatitudes, according
  to ref. \cite{afgl:1986}. A cirrus cloud at an altitude of 10 km
  has been added; it is 1 km deep, with a liquid water content of $0.06
gm^{-3}$ and an effective droplet radius of $50\mu m$
  The source irradiance is chosen in accordance to ref. \cite{kato:1999}.
  This atmosphere is considered aerosol-free}
  \label{transmissivity_clouds}
\end{flushleft}
\end{figure}

In order to study the possibility of establish a quantum
communication channel, the presence of clouds has to be considered
as well. In order to do this, we have added clouds to the atmosphere
\cite{afgl:1986} without aerosols. We set at an altitude of 10 km, a
1 km deep layer of clouds, whose liquid water content is $0.06
gm^{-3}$ and the effective droplet radius is $50\mu m$. This
configuration matches a cirrus and the estimation of direct downward
transmittance is depicted in the figure
(\ref{transmissivity_clouds}).

\begin{figure}[h]
\begin{flushleft}
  \includegraphics[width=8.5cm]{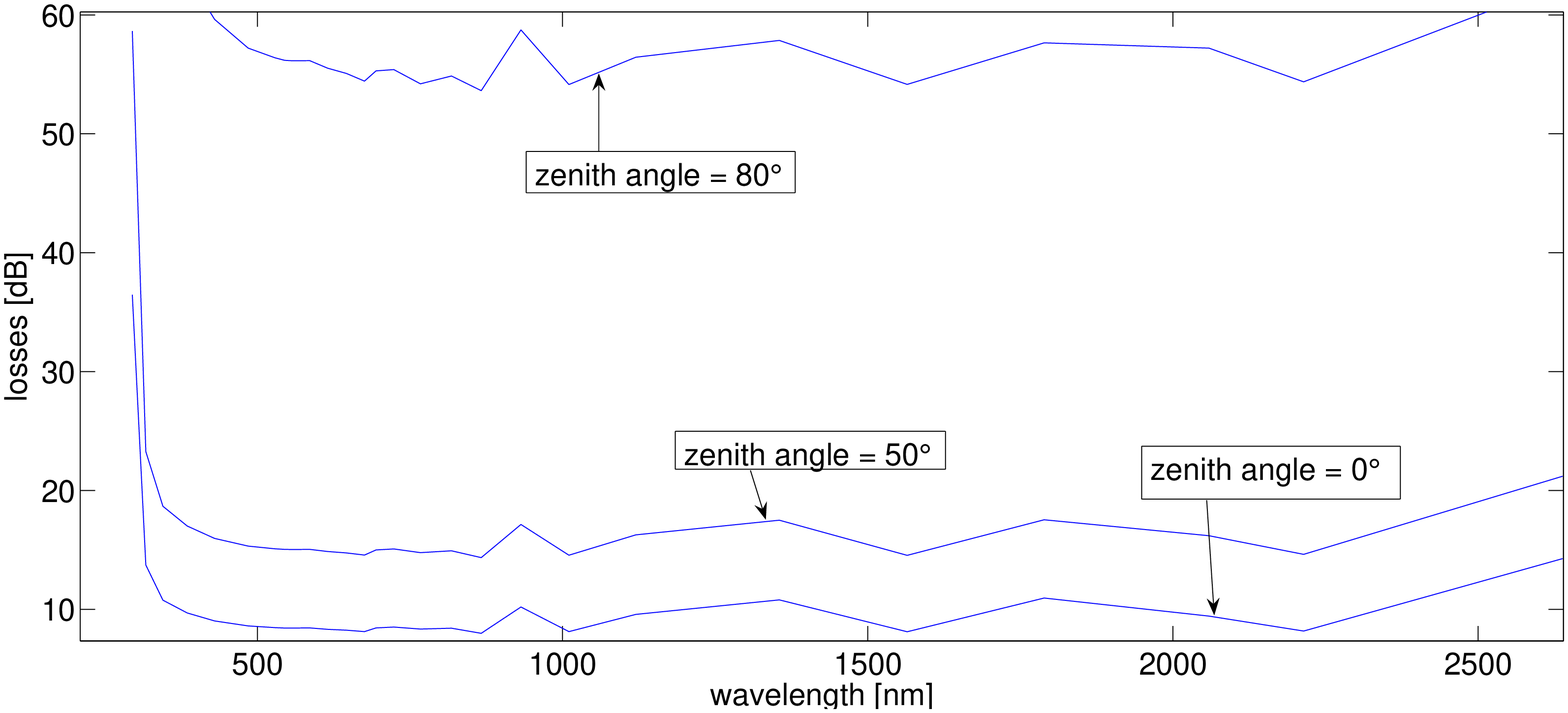}\\
  \caption{Losses in dB vs wavelength in nm for 0$^o$, 50$^o$ and 80$^o$ source zenith angles, in the range
  between 256.3 nm and 2638.5 nm.
  The atmosphere is in summer conditions and at midlatitudes, according
  to ref. \cite{afgl:1986}. A cirrus cloud at an altitude of 10 km
  has been added; it is 1 km deep, with a liquid water content of $0.06
gm^{-3}$ and an effective droplet radius of $50\mu m$
  The source irradiance is chosen in accordance to ref. \cite{kato:1999}.
  This atmosphere is considered aerosol-free}
  \label{losses_clouds}
\end{flushleft}
\end{figure}

Liquid water content and effective droplet radius are translated
into optical properties in \cite{hu_st}. As it can be seen in the
figure (\ref{losses_clouds}), the presence of this kind of cloud
doesn't disable the communication. For 0$^o$\ and 50$^o$\ zenith
angles, the losses are smaller than 60 dB starting from the 295.1 nm
wavelength, nevertheless they remain always around 15-20dB. At the
zenith angle of 80$^o$, losses are dramatically close to 60 dB.
Thus, these results suggest that even the presence of thin clouds as
cirri makes the transmission substantially delicate. On the other
hand the presence of any other kind of clouds with a higher water
content (stratus $L=0.28 gm^{-3}$, cumulus $L=0.26 gm^{-3}$,
cumulonimbus $L=1 gm^{-3}$, stratocumulus $L=0.44 gm^{-3}$, ...)
makes the communication impossible. This information together with a
description of passages of different clouds within different
perturbations and statistical data on average meteorological
evolution in a year for  a given station  allows an estimate of the
available time for  transmission from a specific place.

The program allows a similar analysis for  fog with different
degrees of optical depth as well.

\subsection{Comparison between two extremely different conditions}
\label{subsec:diff_cond}

\begin{figure}[h]
\begin{flushleft}
  \includegraphics[width=8.5cm]{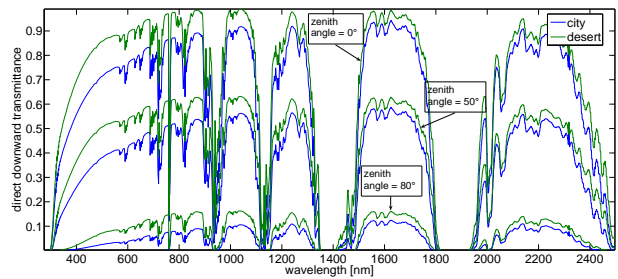}\\
  \caption{Direct downward transmittance (ratio between the direct downward
  irradiance at the Earth surface with respect to the source irradiance)
  vs wavelength in nm for 0$^o$, 50$^o$ and 80$^o$ source zenith angles, in the range
  between 256.3 nm and 2638.5 nm. The atmosphere is in summer
  conditions and at midlatitudes, according to ref. \cite{afgl:1986}.
  The source irradiance is chosen in accordance to ref. \cite{kato:1999}.
  In the first case, there is a urban aerosol environment and a 90\%
  relative humidity and on the other side, there is a dry desert without aerosols}
  \label{transmissivity_citydesert}
\end{flushleft}
\end{figure}

Then we want to get an idea of how is transmissivity for two
extremely different conditions. On one side there is a city
environment with relevant aerosols concentrations and 90\% relative
humidity, on the other side a dry desert without aerosols. The
results for these two cases are depicted in figure
(\ref{transmissivity_citydesert}) for source zenith angle of 0$^o$,
50$^o$ and 80$^o$.

As we could expect, a dry desert is a much better environment for
quantum communication than a humid city. The losses in the desert
are always at least 3 dB smaller than in the city\\
Anyway, either a dry desert and a city are secure environment for
quantum communications. For 0$^o$\ and 50$^o$\ zenith angles,
starting with wavelength equal to 295.1 nm, losses are lower than 60
dB, for 80$^o$\ the first secure wavelength is 317.3 nm.\\
We want to point out once more that the losses calculated so far,
are only expect from the atmosphere. The security limit of 60 dB is
a total loss limit, including, for instance, quantum efficiency of
detectors, optical losses in the devices,...\\
For all the cases under consideration, the best range for
telecommunications is from 700 nm to 900 nm.

\begin{figure}[h]
\begin{flushleft}
  \includegraphics[width=8.5cm]{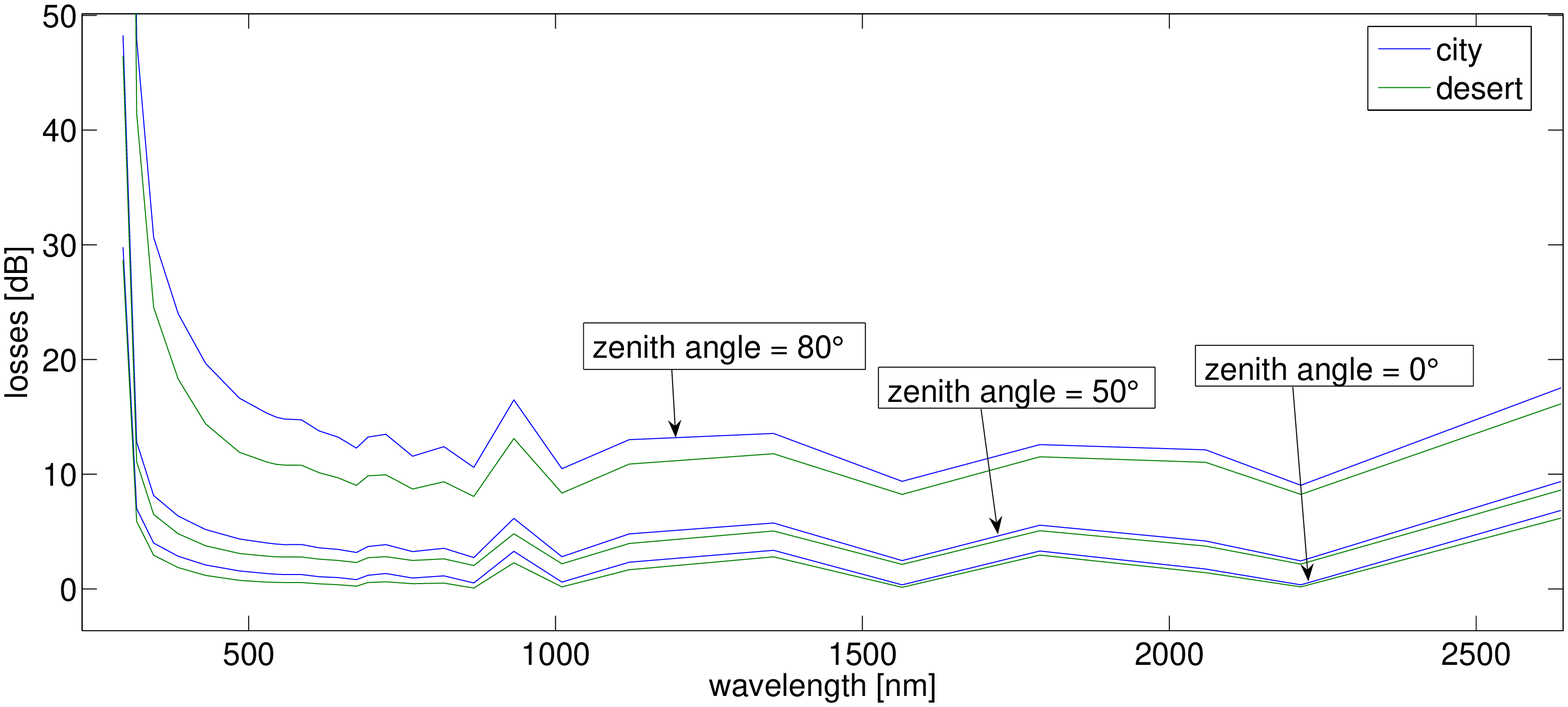}\\
  \caption{Losses in dB vs wavelength in nm for 0$^o$, 50$^o$ and 80$^o$ source zenith angles, in the range
  between 256.3 nm and 2638.5 nm. The atmosphere is in summer
  conditions and at midlatitudes, according to ref. \cite{afgl:1986}.
  The source irradiance is chosen in accordance to ref. \cite{kato:1999}.
  In the first case, there is a urban aerosol environment and a 90\%
  relative humidity and on the other side, there is a dry desert without aerosols}
  \label{losses_citydesert}
\end{flushleft}
\end{figure}
\subsection{Quantum Entanglement over the Danube}
\label{sec:quant_ent_over_danube}

Finally we would like to consider the realistic situation of some
experiment.

As a first example, we consider a quantum entanglement distribution
experiment \cite{danube} performed over the Danube in Vienna, for a
distance of 600m. The information we can infer from the paper some
of the meteorological conditions of when such experiment was
realized, e.g. the temperature was around 0$^o$C and  wind had
strength up to 50 km/h. The bottom (at the Danube level) of the
atmosphere is in summer conditions and at midlatitudes, according to
ref. \cite{afgl:1986}. Urban aerosols are used. The source
irradiance is chosen in accordance to ref. \cite{kato:1999}.

\begin{figure}[h]
\begin{flushleft}
  \includegraphics[width=8.5cm]{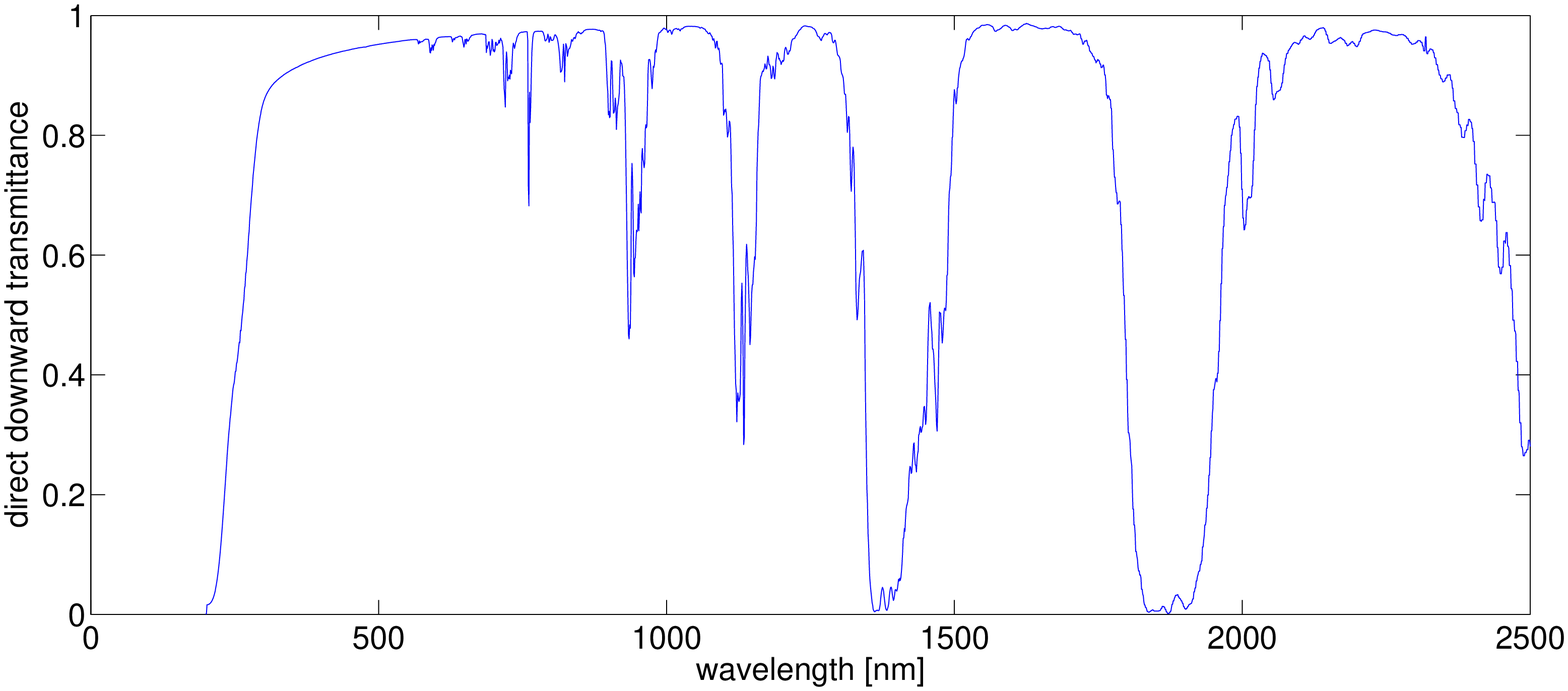}\\
  \caption{Direct downward transmittance (ratio between the direct downward
  irradiance at the Earth surface with respect to the source irradiance)
  vs wavelength in nm, in the range between 256.3 nm and 2638.5 nm. The bottom (at the Danube level)
  of the atmosphere is in summer conditions and at midlatitudes, according to ref.
  \cite{afgl:1986}. Urban aerosols are used.
  The source irradiance is chosen in accordance to ref. \cite{kato:1999}.
  The path the light has to cross is 600 m long, the wind blows at 50 km/h and
  the surface temperature is 0$^o$C}
  \label{transmittance_danube}
\end{flushleft}
\end{figure}

As an application of our program to a realistic situation, here we
report the atmospheric effects for this experiment as deduced from
our analysis. Although the receivers were located at a distance of
either 150 m and 500 m from the source of entangled photon, we
discuss the atmospheric effects over 600 m, the distance between the
two  receivers. The direct downward transmittance is shown in figure
(\ref{transmittance_danube}).

\begin{figure}[h]
\begin{flushleft}
  \includegraphics[width=8.5cm]{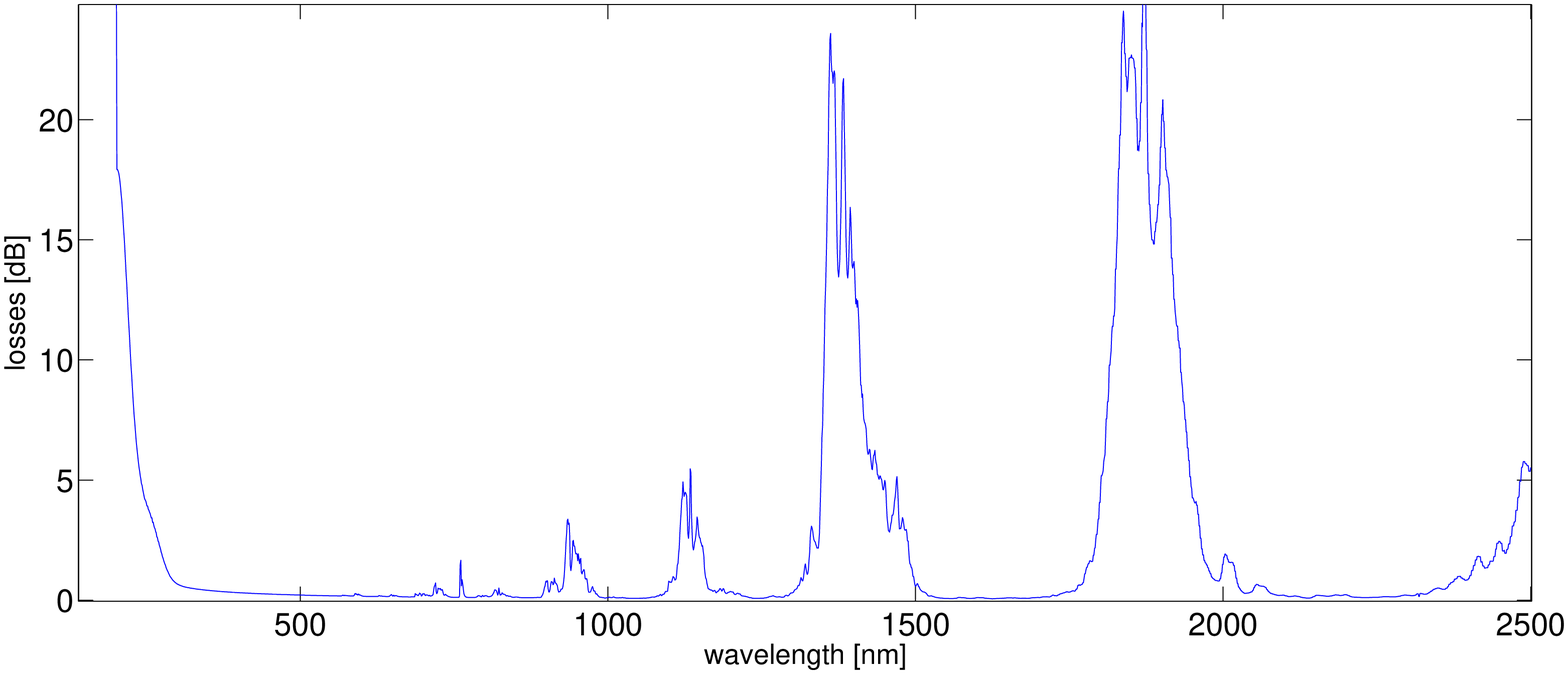}\\
  \caption{Losses in dB vs wavelength in nm, in the range between 256 nm and 2638 nm. The bottom (at the Danube level)
  of the atmosphere is in summer conditions and at midlatitudes, according to ref.
  \cite{afgl:1986}. Urban aerosols are used.
  The source irradiance is chosen in accordance to ref. \cite{kato:1999}.
  The path the light has to cross is 600 m long, the wind blows at 50 km/h and
  the surface temperature is 0$^o$C. The best range is, as usual so far, from 700 nm to 900 nm}
  \label{losses_danube}
\end{flushleft}
\end{figure}

For 810 nm (the wavelength in the experiment), the direct downward
transmittance percentage is about 94\%. In \cite{danube} it is
reported that "The attenuation in each of the links was about 12dB",
but no more indications are given about the sources of attenuation.
According to our simulation, the atmospheric losses at that
wavelength are less than 0.3 dB (see fig.\ref{losses_danube}). We
can thus suppose that the main attenuation factors were the optical

losses and finite quantum efficiency of detectors.

\subsection{144 km transmission}
\label{subsec:120km}

As hinted in the Introduction, an ongoing experiment at Canary
Islands is devoted to establish a 140 km quantum-link. Here we
discuss photon transmission for a reasonable range of atmospheric
conditions in a foreseen Canary Island scenario.

Generic environmental conditions are taken into considerations: the
atmosphere is in summer conditions and at midlatitudes, according to
ref. \cite{afgl:1986}. Maritime aerosols are used. The source
irradiance is chosen in accordance to ref. \cite{kato:1999}, a 2mm
monthly averaged water precipitation (value that will scale the
water vapor profile accordingly).

\begin{figure}[h]
\begin{flushleft}
  \includegraphics[width=8.5cm]{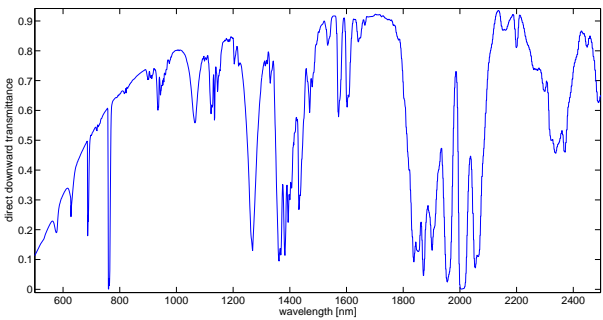}\\
  \caption{Direct downward transmittance (ratio between the direct downward
  irradiance at the Earth surface with respect to the source irradiance)
  vs wavelength in nm, in the range between 256.3 nm and 2638.5 nm.
  The scenario is at the Canary islands. The atmosphere is in summer conditions and at midlatitudes, according to ref.
  \cite{afgl:1986}. Maritime aerosols are used.
  The source irradiance is chosen in accordance to ref. \cite{kato:1999}.
  The path the light has to cross is 144 km long, the average monthly precipitation is 2
  mm}
  \label{transmittance_laspalmas}
\end{flushleft}
\end{figure}

The results of a transmission on a 144 km distance in this scenario
are reported in figure (\ref{transmittance_laspalmas}).

\begin{figure}[h]
\begin{flushleft}
  \includegraphics[width=8.5cm]{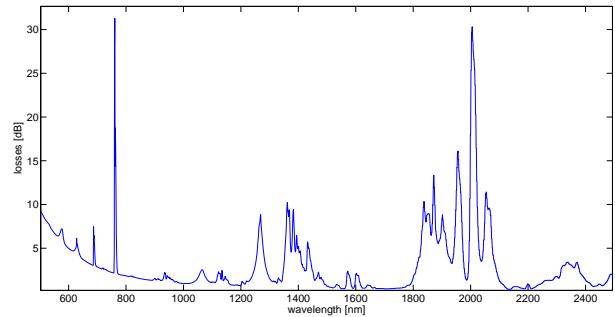}\\
  \caption{Losses in dB vs wavelength in nm, in the range between 256.3 nm and 2638.5 nm.
  The scenario is at the Canary islands. The atmosphere is in summer conditions and at midlatitudes, according to ref.
  \cite{afgl:1986}. Maritime aerosols are used.
  The source irradiance is chosen in accordance to ref. \cite{kato:1999}.
  The path the light has to cross is 120 km long, the average monthly precipitation is 2
  mm}
  \label{losses_laspalmas}
\end{flushleft}
\end{figure}

Results show that higher transmission percentages can be obtained
for high  wavelengths. Anyway, the losses are always less than 20 dB
in the range from 700 nm to 900 nm.\\
The secure communication (losses lower than 60 dB) starts from the
wavelength equal to 345.1 nm (see figure (\ref{losses_laspalmas})).

From a comparison of figures \ref{losses_laspalmas} and
\ref{losses_danube} one can also appreciate as different atmospheric
conditions (aerosols, humidity, etc.) affect specific absorption
regions.

\section{Conclusions}
\label{sec:conclusions}

In this paper we have presented some preliminary results on
atmospheric interaction with photons, obtained by
using the free source library libRadtran.\\
Our results show that  a secure communication can be established
under many realistic meteorological conditions  even up to only
$10^o$ from horizon. Thus, a Earth-satellite quantum channel can be
realized for a large fraction of visibility each orbit.

Furthermore, our results can be used for a first estimate of the
fraction of time per year when a secure communication quantum
channel with a certain satellite can be achieved.

A further deeper analysis of atmospheric effects based on this
approach could effectively be a useful tool for predicting precisely
the performances of a quantum communication channel in various
realistic operative meteorological situations.

\section{Acknowledgements}
This work has been supported by Regione Piemonte (E14), by MIUR FIRB
RBAU01L5AZ-002 and by "San Paolo foundation".

\end{document}